# Human Rights for the Digital Age: A Comprehensive Review

Shaleeza Yaqoob Siddiqui
*Software Engineering*
*Gift University*
Gujranwala, Pakistan
201400053@gift.edu.pk

Sara Farooqi
*Software Engineering*
*Gift University*
Gujranwala, Pakistan
201400023@gift.edu.pk

Wajeeh ur Rehman
*Software Engineering*
*Gift University*
Gujranwala, Pakistan
211400021@gift.edu.pk

Laiba Zulfiqar
*Software Engineering*
*Gift University*
Gujranwala, Pakistan
211400113@gift.edu.pk

*Abstract*— The emergence of digital technology has fundamentally transformed all facets of human existence, posing important queries about the safeguarding and implementation of human rights in the digital domain. The research focuses on important topics including privacy, freedom of speech, and information access. The methodology involves an extensive review of existing literature, legal frameworks, and relevant case studies to provide a comprehensive understanding of the intersection between technology and human rights. The paper highlights the challenges posed by surveillance, data breaches, and the digital divide while also exploring the role of international law and policy in safeguarding digital rights. The review highlights the significance of modifying human rights frameworks for the digital era, pointing out gaps in existing research and offering recommendations for future investigations.

## I. INTRODUCTION

The article explores the relationship between human rights and the digital world, specifically whether having access to the Internet should be considered a human right. Kay Mathiesen challenges Vinton Cerf's belief that human rights are universal and not connected to specific technologies. She argues that new rights should be recognized as technology evolves, like the Internet, because they help protect basic rights like free speech and access to information. Mathiesen believes that we need a "Declaration of Digital Rights" to address how modern technology affects traditional human rights. While she agrees that Cerf's concerns about calling Internet access a human right are valid, she emphasizes the need to adapt human rights to fit today's realities. The article also points out that important human rights documents already stress the importance of access to information, which in the digital age is both enhanced and challenged by new technology, making it crucial to protect these rights for everyone.

## II. PRIVACY IN THE DIGITAL AGE

The digital era has brought about significant challenges to the fundamental human right of privacy, which is acknowledged in various international legal frameworks. The traditional definition of privacy as the freedom from unwarranted intrusion and the ability to manage personal information which has been complicated by the rise of digital technologies. As public and private spaces become more indistinct, the understanding of privacy is changing. In the digital age, vast amounts of personal data are collected, stored, and analyzed, often without the individual's explicit knowledge or consent.

## III. FREEDOM OF EXPRESSION

In the digital age, freedom of expression has been transformed by platforms like social media, blogs, and forums, enabling global communication and breaking down traditional barriers. However, this connectivity also introduces challenges, particularly in censorship. Both governments and private companies regulate online content to maintain order, protect security, or prevent harm. Governments may impose strict internet controls or shut down networks during unrest, while private companies enforce content guidelines that can lead to the removal of posts or account suspensions. This creates a complex balance between expanding free speech and managing harmful content, highlighting the ongoing tension in digital communication.

## IV. THE IMPACT OF THE DIGITAL ERA ON HUMAN RIGHTS

The emergence of the digital era has resulted in a fundamental change in the perception and application of human rights. People now have more ways than ever to express themselves, get information, and take part in democracy thanks to the widespread use of digital technologies. On the other hand, worries about mass surveillance, the dissemination of false information, and the erosion of privacy have also been sparked by this increased connectivity. Researchers studying the effects of the digital era on human rights have started to make the case that internet access is a basic human right in and of itself (Klang & Murray, 2005). Others have issued warnings about the possible dangers of the internet, such as the propagation of hate speech and the maintenance of current social injustices (Benkler, 2006). Human rights are closely related to digital technologies in a complex and subtle way. On the one hand, people can now exercise their right to free speech and obtain information on a never-before-seen scale thanks to digital technologies. However, the growing reliance on digital technologies has also brought about new risks to human rights, such as the possibility of widespread surveillance and a decline in privacy (Solove, 2004).

## V. DISCUSSION

This article makes the case that human rights offer a crucial framework for comprehending how digital technologies affect people individually and as a society. By analyzing the connection between digital technologies and human rights, we can gain a deeper comprehension of how digital technologies are influencing our conception of human rights and how human rights frameworks can be modified to meet the demands of these emerging technologies. A major issue raised by digital technologies is striking a balance between the rights of the individual and the interests of the group. For instance, the necessity to shield people from hate speech and other harmful content must be weighed against the right to freedom of expression. Likewise, the necessity to protect public safety and deter crime must be weighed against the right to privacy. It is crucial to create new frameworks and procedures for protecting human rights in the digital era in order to address these issues. This could entail creating new approaches to human rights that take into consideration the rapidly changing nature of digital technologies, as well as modifying current human rights frameworks to address the particular difficulties presented by these technologies. In the end, academics, decision-makers in government, and business executives will need to put up a consistent effort to safeguard human rights in the digital era. We can make sure that digital technologies are utilized to uphold rather than diminish human rights and well-being by cooperating to create new frameworks and procedures for doing so.

## VI. CONCLUSION

The digital era has fundamentally reshaped the landscape of human rights, offering unprecedented opportunities for expression, access to information, and democratic participation. However, these advancements have come with significant challenges, including the threats of mass surveillance, privacy erosion, and the spread of misinformation. As digital technologies continue to evolve, the relationship between human rights and these technologies remains complex and multifaceted. It is clear that while digital tools can empower individuals and enhance freedoms, they also pose substantial risks that must be addressed. Protecting human rights in the digital age requires a delicate balance between embracing technological innovation and safeguarding fundamental freedoms. This balance can be achieved through the development of robust legal frameworks, international cooperation, and a commitment to ethical standards in the digital realm. As we move forward, it is crucial that policymakers, technologists, and civil society work together to ensure that the benefits of digital technologies are realized without compromising the essential rights and freedoms that underpin human dignity. By doing so, we can create a digital future that respects and upholds human rights for all.